\documentclass[10pt]{article} 
\usepackage{pslatex} 
\usepackage[paperheight=22.86cm,paperwidth=15.24cm,left=1.27cm,right=1.27cm,top=1.6cm,bottom=1.6cm,headheight=1in]{geometry}
\usepackage[utf8]{inputenc}
\usepackage[T1]{fontenc}
\usepackage[hidelinks]{hyperref}
\usepackage{graphicx}
\usepackage{enumerate}
\usepackage[font={small,sf},format=plain,labelfont=bf,up]{caption} 
\usepackage{sectsty}
\sectionfont{\fontsize{12pt}{12pt}\selectfont} 
\usepackage{titlesec}
\titlelabel{\thetitle.\quad}

\usepackage{caption}
\usepackage{subfig}
\usepackage{enumitem}
\usepackage{hyperref}
\usepackage{multirow}
\usepackage{url}
\usepackage{float}
\usepackage[numbers]{natbib}
\begin{document}

\begin{center}
    {\fontsize{14pt}{14pt}\selectfont \textbf{Quantifying Susceptibility to Spear Phishing in a \\High School Environment Using Signal Detection Theory}\par}
    \vspace{\baselineskip}
    \vspace{\baselineskip}
    ${~P. Unchit^1, ~S. Das^{1,2}, ~A. Kim^1, ~and ~L. J Camp^1}$
    \vspace{\baselineskip}
    \\
    {\fontsize{10pt}{10pt}\selectfont 1. Indiana University Bloomington, ~2. University of Denver\par}
    {\fontsize{10pt}{10pt}\selectfont (punchit, sancdas, anykim, ljcamp) @iu.edu\par}
\end{center}
\linespread{0.7}
\renewcommand*{\bibfont}{\small}
\section*{Abstract}
\vspace{-3mm}
{\fontsize{9pt}{9pt}\selectfont
\linespread{0.2}
{
Spear phishing is a deceptive attack that uses social engineering to obtain confidential information through targeted victimization. It is distinguished by its use of social cues and personalized information to target specific victims. Previous work on resilience to spear phishing has focused on convenience samples, with a disproportionate focus on students. In contrast, here, we report on an evaluation of a high school community. We engaged $57$ high school students and faculty members ($12$ high school students, $45$ staff members) as participants in research utilizing signal detection theory (SDT). Through scenario-based analysis, participants tasked with distinguishing phishing emails from authentic emails. The results revealed an overconfidence bias in self-detection from the participants, regardless of their technical background. These findings are critical for evaluating the decision-making of underrepresented populations and protecting people from potential spear phishing attacks by examining human susceptibility.}\par}
\section*{Keywords}
\vspace{-2mm}
{\fontsize{10pt}{10pt}\selectfont Phishing, Spear Phishing, High School, User Study, Usable security\par}

\section{Introduction}
\vspace{-2mm}
Phishing is used to obtain confidential information, install malware, obtain funds, or steal resources~\cite{hadnagy2010social}. Targeted phishing is a critical component of that; for example, phishing attacks on Zoom increased four orders of magnitude between March and April 2020 and COVID-19-related phishing, including misinformation as well as attacks on the benefits for the newly unemployed. The most targeted form of phishing attack is~\textit{spear phishing}~\cite{apwg}. As spear phishing is a challenge essentially grounded in human behavior and decision-making~\cite{pattinson2012some}, solutions should be informed by human subject evaluations as well.

Conversely, studies on phishing show a bias toward machine learning and purely technical solutions, with only $13.9\%$ of published papers on phishing in the ACM Digital Library utilizing human participants or user-centered methodologies~\cite{das2019all}. Even when research does involve human subjects, it often studies convenience samples, specifically university students. Investigating high school students is particularly important, as previous research has shown that age is a critical factor in predicting susceptibility to phishing attacks~\cite{kumaraguru2009school,lastdrager2017effective,nicholson2020investigating}. Improved understanding of participants' mindsets when they click on a malicious email link can enable robust defensive and offensive techniques against spear phishing attacks. In order to contribute to this understanding, we combined phishing detection with signal detection theory (SDT) to explore how spear phishing cues impact this population~\cite{canfield2016quantifying}. SDT is often used to effectively measure and differentiate between present patterns and figuratively noisy distractions~\cite{martin2018signal}.

Specifically, we conducted a user study focusing on $57$ high school students and staff members to explore the less-observed correlation between participant mentalities and email spear phishing attacks. Our goal was to address the following research questions:
\begin{itemize}
    \item {RQ1: How confident are participants in distinguishing between legitimate and non-legitimate spear phishing content over email?}
    \item{RQ2: How does age affect a user's ability to distinguish between legitimate and non-legitimate spear phishing content over email?}
\end{itemize}

\section{Related Work}
\label{related}
\vspace{-2mm}
The U.S. Department of Homeland Security identified the sequence of actions taken to craft a spear phishing attack: (1) identify the target, (2) meticulously craft the message with the intent of the recipient taking immediate action, and (3) deliver the message from a counterfeit email address~\cite{phishing2018phooled}. Rajivan et al. found that phishing emails with \lq\lq specific attack strategies (e.g., sending notifications, use of authoritative tone, or expressing shared interest)\rq\rq~were found to be more successful~\cite{rajivan2018creative}. The use of social engineering through psychological manipulation can establish trust, and, as a result, lure in victims~\cite{hatfield2018social}.

Previous research on phishing has focused on software- or hardware-based solutions, such as toolbars, machine learning models, and warning indicators~\cite{das2020user}. Although significant advances in technology-based tools have emerged~\cite{prakash2010phishnet,wu2006security, xiang2011cantina+}, less research has focused on end users~\cite{das2019all}. Yet, the need for such research has long been recognized; in 2008, Friedrichs et al. argued that humans must be studied to stop web-based identity theft, including phishing attacks \cite{friedrichs2008threat}. Such insights become even more important in light of Karakasiliotis et al.'s findings that only 36\% of their study's participants could identify legitimate websites. Only 45\% of participants could correctly identify malicious websites~\cite{karakasiliotis2006assessing}. Dhamija et al. found that visual deception can fool even sophisticated users; a good phishing website fooled 90\% of the participants in their study~\cite{dhamija2006phishing}. Fewer studies have focused on more vulnerable populations, such as younger students. In our background research, we did not find any studies focused on high school students or staff. Thus, we specifically selected a high school environment for our study.

In 2016, Canfield et al. performed two experiments comparing detection and performance using SDT. They found that \lq\lq Greater sensitivity was positively correlated with confidence. Greater willingness to treat emails as legitimate was negatively correlated with perceived consequences from their actions and positively correlated with confidence\rq\rq~\cite{canfield2016quantifying}. We implemented SDT in our research by analyzing the \lq stimulus,\rq~which triggers the decision-making in users. To evaluate the efficacy of the stimulus, we measured hits, misses, false alarms, and correct rejections (i.e., true positive, false negative, false positive, and true negative). We analyzed how users chose to click or not click links sent via electronic mail. The use of SDT enabled us to evaluate which sections of the phishing email arouse suspicion when they are present~\cite{canfield2016quantifying}.
\section{Methodology}
\vspace{-2mm}
To explore the relationship between the phishing susceptibility of high school students and their educators, we wanted to see what email cues both groups notice when deciding to click (or not click) on a malicious link. We conducted a non-experimental, quantitative correlation analysis by collecting data through a descriptive survey to check phishing susceptibility outcomes, age differences, and confidence levels. We primarily collected data from high school students and staff at a suburban high school in the United States. We obtained approval from the Ethical Review Board before beginning this experiment.

\subsection{Recruitment}
\vspace{-2mm}
To begin, we instituted a collaboration with a suburban high school from the Midwestern part of the United States.  As most high school students were under the age of 18, parental permission was required on a paper version of an informed consent document. We only allowed people to participate after their form was signed and approved by the staff and the students' parents. During the recruitment phase, we engaged with language arts classrooms to find willing research participants. English language arts classes were chosen because all students were required to enroll in these classes to graduate. The study was also advertised to every student in the building during the morning school announcements. We also distributed flyers advertising the study to $200$ participants. Students who turned in the paper consent forms then received emails that contained an electronic form of the survey. To recruit teachers and faculty members, we sent out emails containing the link to the consent form and questionnaire. Because the study was announced beforehand, teachers and faculty were expecting this recruitment email. The participants received an incentive at the end of the survey by choosing to enter a drawing for Starbucks gift cards. Our power analysis showed that we required sample size of more than $50$ participants. We obtained a complete response set from $57$ participants in our final data set. 

\subsection{Survey Instrument and Study Design}
\vspace{-2mm}
The survey consisted of three parts: the informed consent information, the demographic questionnaire, and the actual phishing susceptibility assessment. We utilized Google Forms as the tool to provide the survey questionnaire because it was easily accessible to both students and teachers. The first author anonymized the data so that personally identifiable information would not be shared with anyone else, including other researchers. Participants began by opening a Google Forms link from their email and confirming their status as a student or a staff member of the high school. The staff needed to confirm their consent to the study, while students would move on to the next step due to their parents having already agreed via the consent form. Next, participants answered a set of demographic questions regarding their age group (and not their specific date of birth to reduce the risk of disclosure of identifiable information). Afterward, the participants were presented with ten questions to assess their spear phishing susceptibility through the use of images of phishing emails. We selected images instead of asking them to go through actual emails to mitigate any concern that they may respond to malicious messages. The participants classified the images as \lq\lq regular email\rq\rq~or \lq\lq phishing email\rq\rq. For each question, the participants rated their confidence in their decision, from least to most confident using a five-point Likert scale.

\begin{figure}[h]
    \centering
\begin{minipage}{0.45\textwidth}
    \includegraphics[width=6cm,height=4cm]{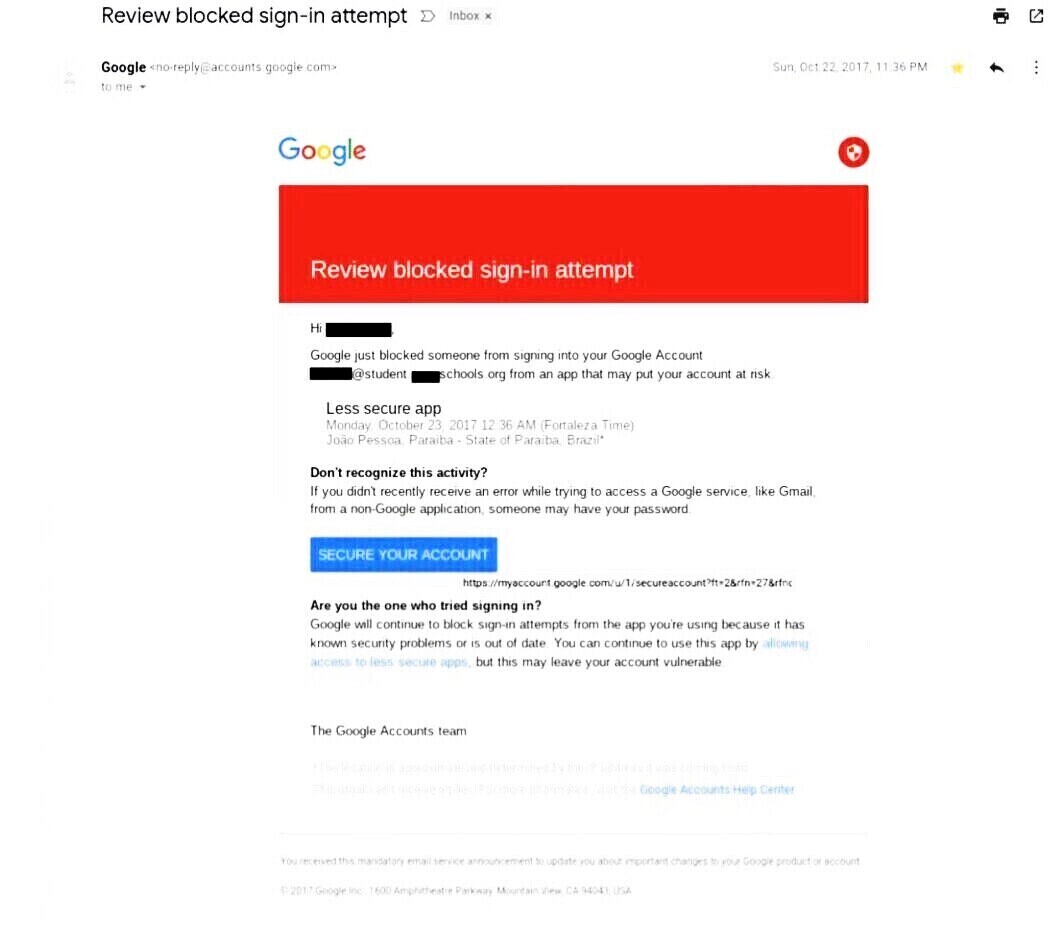}
    \caption{Example of an Authentic Email Displayed to the Participants in the Survey}
    \label{fig:auth}
\end{minipage}
\hspace{1cm}
\begin{minipage}{0.45\textwidth}
    \includegraphics[width=6cm,height=4cm]{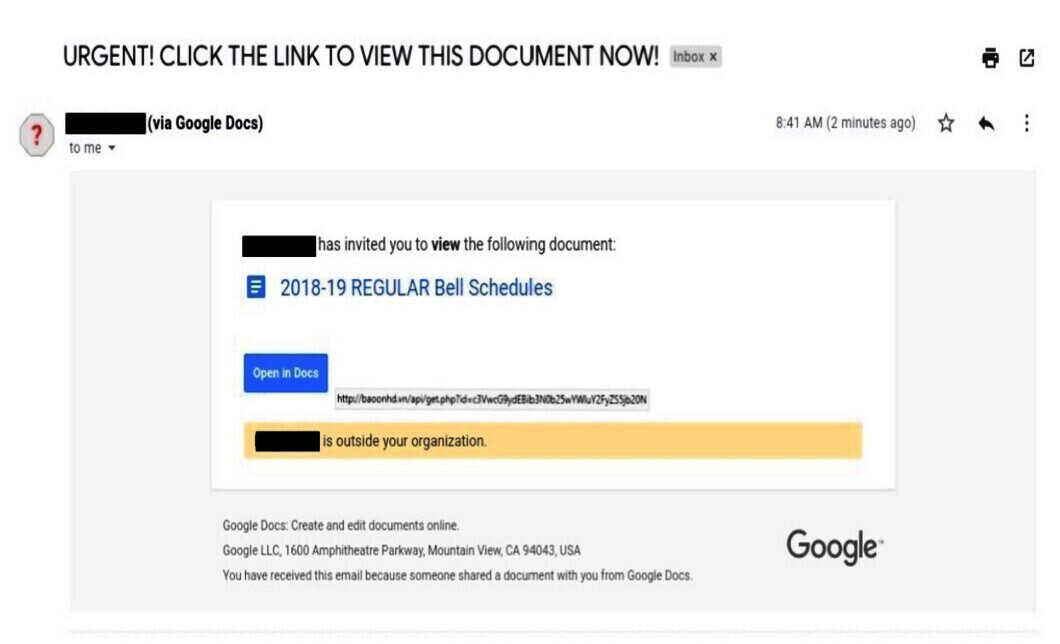}
    \caption{Example of a Phishing Email Displayed to the Participants in the Survey}
    \label{fig:phish}
\end{minipage}
\end{figure}

\textbf{Spear Phishing Susceptibility}:
Based on prior phishing research, there are three main factors identified in most phishing emails: anonymous senders, suspicious URLs or installations, and a sense of urgency~\cite{fette2007learning}. Figure~\ref{fig:phish} is an example that shows the present signs of a harmful phishing email such as: an anonymous sender (e.g., \lq\lq is outside your organization\rq\rq), a sense of urgency (e.g., \lq\lq URGENT! CLICK THE LINK\rq\rq), a suspicious URL (e.g., \lq\lq http://baoonhd.vn/api/get.php?...\rq\rq), and a risky action (e.g., clicking on \lq\lq Open in Docs\rq\rq). In contrast, Figure~\ref{fig:auth} shows an authentic email from Google, as seen by the trustworthy email address, the accurate website link, and the valid email format. Non-phishing examples were adopted from personal school emails that the high school staff and students received earlier, and at least one individual reported as suspicious. This data was obtained from the high school staff and IT support, who anonymized the email samples.

Phishing examples were adopted from the Berkeley Phishing Examples Archive (PEA)~\footnote{\url{https://security.berkeley.edu/education-awareness/phishing/phishing-examples-archive}}. The adopted phishing emails were modified to include the name of the school and actual school activities, including grades and exams. The images were edited to address the participants' real names and roles (teacher or student). Google documents addressed school-specific information to check the participants' susceptibility to spear phishing emails. The signals that were used in the phishing emails were (a) the greeting, (b) suspicious URLs with a deceptive name or IP address, (c) content that did not match the ostensible sender and subject, (d) requests for urgent action, and (e) grammatical or typographical errors. We selected this set of signals based on a 2016 study by canfield et al. that similarly focused on detection theory, albeit using an online survey of people aged 19-59~\cite{canfield2016quantifying}.

\vspace{-4mm}
\subsection{Analysis: Method}
\vspace{-2mm}
Once the data collection was complete, we analyzed the data using RStudio and SPSS Statistics. Using SDT, participants' answers were categorized as four possible outcomes: \textit{hit, miss, false alarm, and correct rejection}. Table~\ref{tab:SDT1} shows the signal detection theory outcomes adjusted to become appropriate for this study. The outcomes from the phishing assessment were analyzed in a one-way analysis of variance (ANOVA) to explore the relationship between the independent variable (age group) and the dependent variables (the number of different outcomes and the average confidence levels). The one-way analysis of variance is used to determine whether there are any statistically significant differences between the means of two or more independent (unrelated) groups~\cite{girden1992anova}. For ANOVA, we usually compare three or more groups. For this study, we divided the data set into seven groups.

\begin{table}[h!]
\centering
 \begin{tabular}{|c|c|c|} 
 \hline
  & Respond \lq\lq Regular Email\rq\rq~& Respond \lq\lq Phishing Email\rq\rq~\\
 \hline
 Phishing Email & Miss & Hit\\ 
 \hline
 Authentic Email & Correct Rejection & False Alarm \\
 \hline
 \end{tabular}
 \caption{Modified Signal Detection Theory Implemented to Evaluate Spear Phishing Susceptibility}
 \vspace{-3mm}
 \label{tab:SDT1}
\end{table}
\section{Findings and Discussions}
\vspace{-2mm}
\label{sec:findings}
Our data collection was done over a period of two months. We collected a complete data set of $57$ subjects, who provided their consent and participated in it. Of these $57$ participants, $12$ were students, and $45$ were staff members of the high school. Eight participants were from 12 to 17 years old; four participants were from 18 to 24 years old; $11$ participants were from 25 to 34 years old; $15$ participants were from 35 to 44 years old; $12$ were from 45 to 54 years old; seven were from 55 to 64 years old. Thus, the participants' ages ranged from 12 to 64 years old. This study aimed to determine if there was a significant difference between the age groups (12--17, 18--24, 25--34, 35--44, 45--54, and 55--64 years old), the email outcomes (hit, miss, correct rejection, false alarm), and the confidence levels (Likert scale one through five ratings) using a ten-item test. Results of the ANOVA test are shown in Table~\ref{tab:SDT2}. A significant difference was noted for the hit or miss email outcomes (F(5, 51) = 2.614, p < .035). The correct rejection, false alarm, and all the different confidence levels had no significant difference between the groups.

\begin{table}[h!]
\centering
 \begin{tabular}{|p{2.2cm}|c|p{1cm}|c|p{1cm}|c|c|}
\hline
& & Sum of Sq & df & Mean Square & F & Sig.\\ 
 \hline
 Hit & Between Groups & 13.634 & 5 & 2.727 & 2.614 & 0.035 \\
 \hline
 & Within Groups & 53.208 & 51 & 1.043 & & \\
 \hline
 & Total & 66.842 & 56 & & &\\
 \hline 
 Miss & Between Groups & 13.634 & 5 & 2.727 & 2.614 & 0.035 \\
  \hline
 & Within Groups & 53.208 & 51 & 1.043 & & \\
  \hline
 & Total & 66.842 & 56 & & &\\
 \hline
 Rejection & Between Groups & 4.111 & 5 & 0.822 & 1.292 & 0.282 \\
 \hline
 & Within Groups & 32.451 & 51 & 0.636 & & \\
 \hline
 & Total & 36.561 & 56 & & &\\
 \hline
 FalseAlarm & Between Groups & 4.111 & 5 & 0.822 & 1.292 & 0.282 \\
 \hline
  & Within Groups & 32.451 & 51 & 0.636 & & \\
  \hline
 & Total & 36.561 & 56 & & &\\
 \hline
 HitConf & Between Groups & 2.156 & 5 & 0.431 & 0.976 & 0.441 \\
 \hline
 & Within Groups & 22.079 & 50 & 0.442 & & \\
 \hline
 & Total & 24.234 & 55 & & & \\
 \hline 
 MissConf & Between Groups & 2.954 & 5 & 0.591 & 1.548 & 0.194 \\
 \hline
 & Within Groups & 17.554 & 46 & 0.382 & & \\
 \hline
 & Total& 20.507 & 51 & & & \\
 \hline
 CorrRejConf & Between Groups & 0.812 & 5 & 0.162 & 0.558 & 0.732 \\
 \hline 
 & Within Groups & 14.854 & 51 & 0.291 & & \\
 \hline
 & Total & 15.667 & 56 & & & \\
 \hline
 FalseAlarmConf & Between Groups & 1.457 & 5 & 0.291 & 0.514 & 0.764 \\
 \hline
 & Within Groups & 23.818 & 42 & 0.567 & & \\
 \hline
 & Total & 25.275 & 47 & & &\\
 \hline
 \end{tabular}
 \caption{ANOVA Results of the Different Signals (Hit, Miss, Correct Rejection, and False Alarm) Between and Within Groups (Divided Based on Age)}
 \label{tab:SDT2}
\end{table}

\begin{figure}[h]
    \centering
\begin{minipage}{0.4\textwidth}
    \includegraphics[width=5cm,height=4.2cm]{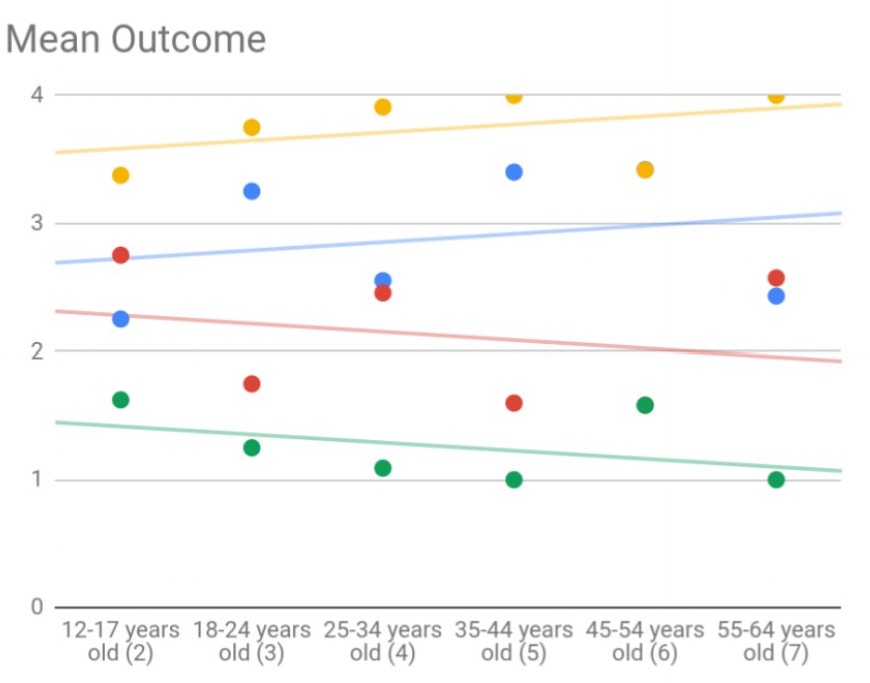}
    \caption{SDT Mean Outcome shows the mean for the email outcomes in a linear transformation from 100\% to a five point scale. It shows (from top to bottom) correct rejection in yellow, correct acceptance (hit) in blue, incorrect acceptance (miss) in red, and false alarm in green.}
    \label{fig:sdt}
\end{minipage}
\hspace{0.5cm}
\begin{minipage}{0.52\textwidth}
\centering
    \includegraphics[width=5.2cm,height=4.3cm]{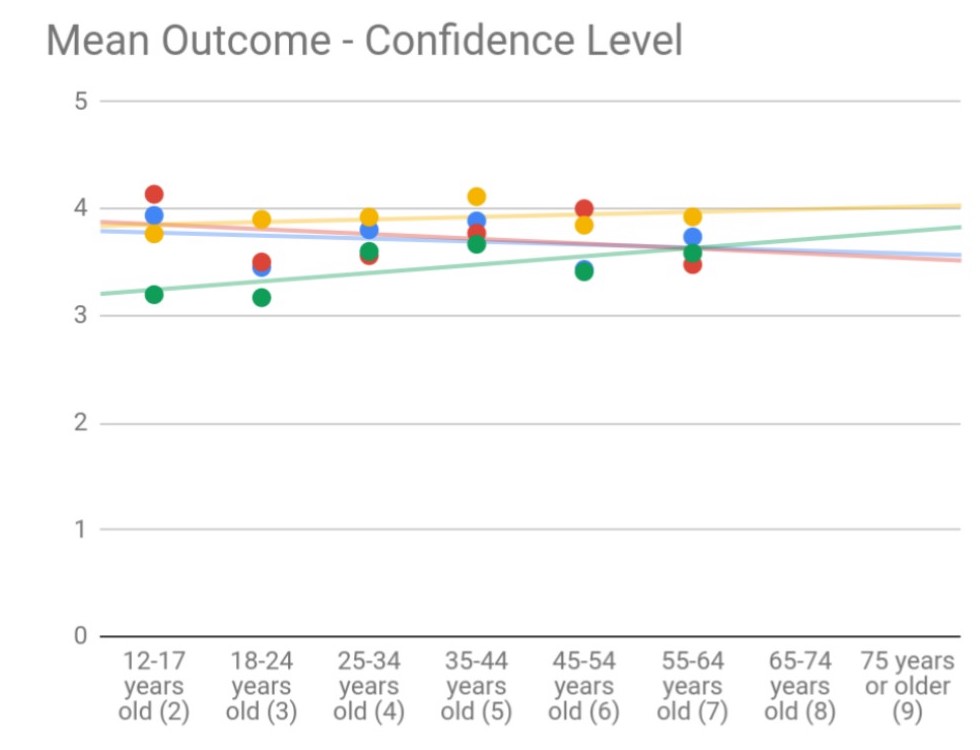}
    \caption{SDT Mean Outcome for Confidence Levels showing the confidence level. Misclassifying phishing email(red) is associated with the same confidence as correct rejection for 12-17 (yellow), with confidence falling with age. False alarm is shown with least confidence in ages 12-17, and increases with with age.}
    \label{fig:sdt2}
    \end{minipage}
\vspace{-2mm}
\end{figure}

The results illustrate a significant number in the hit or miss category, but few correct rejections and false alarms across all the confidence levels. The ANOVA results of the confidence levels of the participants can be seen in Table~\ref{tab:SDT33}. Here, we can say that age plays a significant role in responding to a stimulus, as evidenced by the participants either responding with \lq\lq Authentic Email\rq\rq~or \lq\lq Phishing Email.\rq\rq~
A potential reason for the lack of significance could be that the confidence levels were not precisely represented and that participants' perceived confidence was subjective. One participant's response of a 5 (most confident) could be the same as another participant's 3 (average confidence). Their perceived confidence could also shift throughout the survey; a response of 1 (least confident) could be changed to a 2 (lower confidence) or 3 later on, depending on whether or not the participants believed that the questions were more or less difficult at the beginning of the survey.

Fig~\ref{fig:sdt} shows correct results (yellow, blue) increase with age. Fig~\ref{fig:sdt2} show confidence increasing in false alarms in with age (green), with confidence about correct identification ( and misidentification higher for younger age groups.  Our data revealed that the highest mean for the hit outcome was from age group six (45-54 years old). The second-highest mean for the hit outcome was from age group five (35-44 years old). Groups five and six also had the lowest mean for the miss outcome. In Figure~\ref{fig:sdt}, we show the mean outcome for hit and correct rejection, which has an increasing slope, with a negative correlation with miss and false alarm. Therefore, there is strong evidence that older groups are less susceptible to spear phishing than the younger groups in a high school setting. Figure~\ref{fig:sdt2} shows that the other variables were not significant. This result is quite different from that hypothesized under the \lq digital native\rq~rubrics that argue for younger cohorts' lifetime exposure resulting in improved decision-making (e.g.,~\cite{nikou2019impact}).

\begin{table}[h!]
\centering
 \begin{tabular}{|p{1cm}|c|c|p{1cm}|p{1cm}|p{1cm}|c|c|c|c|} 
 \hline
 & & N & Mean & SD & SE & 95\% CI & 95\% CI & Min & Max \\
 \hline
 & & & & & & LB & UB & & \\
 \hline
 HitConf & 2 & 8 & 3.9375 & 0.78142 & 0.27627 & 3.28 & 4.5908 & 3 & 5 \\
 \hline
 & 3 & 4 & 3.45 & 0.51171 & 0.25586 & 2.6357 & 4.2643 & 2.8 & 4 \\
 \hline
 & 4 & 11 & 3.803 & 0.69848 & 0.2106 & 3.338 & 4.2723 & 3 & 5 \\
 \hline
 & 5 & 15 & 3.8856 & 0.71328 & 0.18417 & 3.4906 & 4.2806 & 2.5 & 5 \\
 \hline
 & 6 & 12 & 3.4306 & 0.40644 & 0.11733 & 3.1723 & 3.6888 & 2.67 & 4 \\
 \hline
 & 7 & 6 & 3.7361 & 0.80003 & 0.32661 & 2.8965 & 4.5757 & 3 & 5 \\
 \hline
 & Total & 56 & 3.7321 & 0.6638 & 0.0887 & 3.5544 & 3.9099 & 2.5 & 5 \\
 \hline
 Miss Conf & 2 & 8 & 4.1354 & 0.42243 & 0.14935 & 3.7823 & 4.4886 & 3.67 & 5 \\
 \hline
 & 3 & 3 & 3.5 & 0.5 & 0.28868 & 2.2579 & 4.7421 & 3 & 4 \\
 \hline
 & 4 & 11 & 3.5606 & 0.57384 & 0.17302 & 3.1751 & 3.9461 & 3 & 4.5 \\
 \hline
 & 5 & 12 & 3.7708 & 0.66962 & 0.1933 & 3.3454 & 4.1963 & 2.5 & 5 \\
 \hline
 & 6 & 11 & 4 & 0.58214 & 0.17552 & 3.6089 & 4.3911 & 3 & 5 \\
 \hline
 & 7 & 7 & 3.4762 & 0.83571 & 0.31587 & 2.7033 & 4.2491 & 2 & 4.33 \\
 \hline
 & Total & 52 & 3.7756 & 0.63412 & 0.08794 & 3.5991 & 3.9522 & 2 & 5 \\
 \hline
 ConRej Conf & 2 & 8 & 3.7646 & 0.65686 & 0.23223 & 3.2154 & 4.3137 & 3 & 5 \\
 \hline
 & 3 & 4 & 3.9 & 0.57991 & 0.28996 & 2.9772 & 4.8228 & 3.33 & 4.67 \\
 \hline
 & 4 & 11 & 3.912 & 0.55523 & 0.16741 & 3.5482 & 4.2942 & 3 & 4.75 \\
 \hline
 & 5 & 15 & 4.1133 & 0.51564 & 0.13314 & 3.8278 & 4.3989 & 3.25 & 5 \\
 \hline
 & 6 & 12 & 3.8458 & 0.55674 & 0.16072 & 3.4921 & 4.1996 & 2.67 & 4.5 \\
 \hline
 & 7 & 7 & 3.9238 & 0.31898 & 0.12056 & 3.6288 & 4.2188 & 3.5 & 4.5 \\
 \hline
 & Total & 57 & 3.9327 & 0.52892 & 0.07006 & 3.7924 & 4.0731 & 2.67 & 5 \\
 \hline
 False Alarm Conf & 2 & 6 & 3.1944 & 0.62731 & 0.2561 & 2.5361 & 3.8528 & 2 & 3.67 \\
 \hline
 & 3 & 3 & 3.1667 & 1.04083 & 0.60093 & 0.5811 & 5.7522 & 2 & 4  \\
 \hline
 & 4 &10 & 3.6 & 0.8756 & 0.27689 & 2.9736 & 4.2264 & 3 & 5 \\
 \hline
 & 5 & 12 & 3.6667 & 0.74874 & 0.21614 & 3.1909 & 4.1424 & 2 & 5 \\
 \hline
 & 6 & 11 & 3.4091 & 0.73547 & 0.22175 & 2.915 & 3.9032 & 2 & 4 \\
 \hline
 & 7 & 6 & 3.5833 & 0.4916 & 0.2069 & 3.0674 & 4.0992 & 3 & 4 \\
 \hline
 & Total & 48 & 3.4931 & 0.73333 & 0.10585 & 3.2801 & 3.706 & 2 & 5 \\
 \hline
 \end{tabular}
 \caption{ANOVA Descriptives for SDT Confidence Levels Outcomes}
 \vspace{-5mm}
 \label{tab:SDT33}
\end{table}

\section{Implications}
\vspace{-2mm}
Spear phishing is an effective form of attack because attackers manipulate their targets, either through luring them in with promises of specific benefits or by coercing them with specific threats~\cite{maurer2011using}. These techniques are designed to lead to impulsive or quick decision-making from the end-users. 
In our findings (Section~\ref{sec:findings}), we leveraged SDT to understand participant decision-making with spear phishing stimuli. When the mean of the outcomes was graphed, the results revealed a positive slope for the hit and correct rejection outcomes, meaning that the older participants tended to be less susceptible to spear phishing. The effects of these relationships can contribute to a better understanding of how people interact with fraudulent acts online. Here we offer recommendations that our findings indicate as ways to increase resilience against spear phishing attacks.

\textbf{Align Anti-Phishing Training with Self-perceived Expertise}: Our work found that older participants were less susceptible to spear phishing than younger participants, as age group six had the highest average number of hits (i.e., correct detection) throughout the experiment. This is aligned with previous research from Sheng et al.~\cite{sheng2010falls}. One reason for this gap may be students' lack of exposure to training geared towards them. For this reason, we recommend introducing phishing training to students at a younger age and aligning it with their self-perceived expertise. Our results show both a high level of incorrect responses and a high level of confidence. This indicates that younger participants may be unaware that they have been the victim of a successful phishing attack.

\textbf{Targeted Risk Communication}: In addition to providing anti-phishing training, organizations should consider providing clear risk communication, especially for younger adults or children. Students may lack an understanding of the technical threats that may be present in their email inbox~\cite{harbach2014using}, believing that they will not be targeted. Thus, the need for context-aware risk communication~\cite{das2020risk} that has been identified as necessary for older adults~\cite{das2019towards,das2020don,garg2012risk} is similarly required for high school student populations.

\textbf{Enable Multi-Factor Authentication}: To create more robust defensive techniques against spear phishing attacks, we need to reduce the risk of compromised credentials. Such compromised credentials can be used to steal sensitive information. Because of this, schools that provide laptops (or require these for online instruction) should consider adopting multi-factor authentication (MFA) for students and staff~\cite{das2018johnny,das2018qualitative,ometov2018multi}. The introduction of these (like other training) should be aligned with user risk mental models~\cite{das2019mfa,das2020mfa,das2019evaluating}.  The issue of over-confidence above also motivates the importance of another factor for authentication (e.g., a hardware token) in addition to their password, which would mitigate the harm of phishing.

\section{Limitations and Future Work}
\vspace{-2mm}
This work, with its focus on the confidence as well as correctness, opens more questions than it answers. Other factors besides age and confidence levels should be studied to gain a holistic understanding of susceptibility to spear phishing. The suburban high school we engaged with has relatively high socio-economic homogeneity, and the study should be repeated with other high schools. To improve diversity, future work should begin with more diverse schools, and then study specific underrepresented populations, such as students with physical or learning disabilities. Interviewing the participants to collect more qualitative data and better understand user decision making is a needed expansion of this work.

\section{Conclusion}
\vspace{-2mm}
With the current rise in spear phishing, especially among vulnerable populations, it is critical to developing tools and educational approaches to train users to differentiate between authentic and malicious emails. To understand spear phishing attack resilience, we studied a population in a high school environment ($N=57$). We found that age and confidence play a critical role in the identification of spear phishing attacks. Our study concludes by providing recommendations for developing anti-phishing training tools and communicating risks and benefits.
 
\section{Acknowledgement}
\vspace{-2mm}
We would like to the participants of the highschool for their valuable contribution, and Stephanie Davis for encouraging the first author throughout the entire data collection process. We would also like to thank Kevin Gingerich from Eli Lilly for their expert advice on phishing and guiding the first author.  This research was supported in part by the National Science Foundation under CNS 1565375, Cisco Research Support, and the Comcast Innovation Fund. Any opinions, findings, and conclusions or recommendations expressed in this material are those of the author(s). They do not necessarily reflect the views of the U.S. Government, NSF, Cisco, Comcast, Indiana U, or the University of Denver.
\section{References}
\vspace{-2mm}
{\fontsize{7pt}{7pt}\selectfont \begingroup
\renewcommand{\section}[2]{}
\bibliographystyle{abbrv}
\bibliography{phishing}

\begin{thebibliography}{10}

\bibitem{apwg}
APWG.
\newblock {\em Phishing Activity Trends Report}, 2020 (accessed June 29, 2020).
\newblock \url{"https://docs.apwg.org/reports/apwg_trends_report_q1_2020.pdf}.

\bibitem{canfield2016quantifying}
C.~I. Canfield, B.~Fischhoff, and A.~Davis.
\newblock {Quantifying Phishing Susceptibility for Detection and Behavior
  Decisions}.
\newblock {\em Human Factors}, 58(8):1158--1172, 2016.

\bibitem{das2020risk}
S.~Das.
\newblock {\em A Risk-reduction-based Incentivization Model for Human-centered
  Multi-factor Authentication}.
\newblock PhD thesis, Indiana University, 2020.

\bibitem{das2020user}
S.~Das, J.~Abbott, S.~Gopavaram, J.~Blythe, and L.~J. Camp.
\newblock User-centered risk communication for safer browsing.
\newblock In {\em First Asia USEC-Workshop on Usable Security, In Conjunction
  with the Twenty-Fourth International Conference International Conference on
  Financial Cryptography and Data Security}, 2020.

\bibitem{das2018johnny}
S.~Das, A.~Dingman, and L.~J. Camp.
\newblock Why johnny doesn’t use two factor a two-phase usability study of
  the fido u2f security key.
\newblock In {\em International Conference on Financial Cryptography and Data
  Security}, pages 160--179. Springer, 2018.

\bibitem{das2019towards}
S.~Das, A.~Kim, B.~Jelen, J.~Streiff, L.~J. Camp, and L.~Huber.
\newblock Towards implementing inclusive authentication technologies for older
  adults.
\newblock {\em Who Are You}, 2019.

\bibitem{das2020don}
S.~Das, A.~Kim, B.~Jelen, J.~Streiff, L.~J. Camp, and L.~Huber.
\newblock Why don’t older adults adopt two-factor authentication?
\newblock {\em Das, S., Kim, A., Jelen, B., Streiff, J., Camp, LJ, \& Huber,
  L.(2020, April). Why Don’t Older Adults Adopt Two-Factor Authentication},
  2020.

\bibitem{das2019all}
S.~Das, A.~Kim, Z.~Tingle, and C.~Nippert-Eng.
\newblock {All About Phishing Exploring User Research through a Systematic
  Literature Review}.
\newblock In {\em 13th International Symposium on Human Aspects of Information
  Security \& Assurance}, 2019.

\bibitem{das2018qualitative}
S.~Das, G.~Russo, A.~C. Dingman, J.~Dev, O.~Kenny, and L.~J. Camp.
\newblock A qualitative study on usability and acceptability of yubico security
  key.
\newblock In {\em 7th Workshop on Socio-Technical Aspects in Security and
  Trust}, pages 28--39, 2018.

\bibitem{das2019mfa}
S.~Das, B.~Wang, and L.~J. Camp.
\newblock {MFA is a Waste of Time! Understanding Negative Connotation Towards
  MFA Applications via User Generated Content}.
\newblock In {\em 13th International Symposium on Human Aspects of Information
  Security \& Assurance (HAISA 2019)}, 2019.

\bibitem{das2020mfa}
S.~Das, B.~Wang, A.~Kim, and L.~J. Camp.
\newblock Mfa is a necessary chore!: Exploring user mental models of
  multi-factor authentication technologies.
\newblock In {\em 53rd Hawaii International Conference on System Sciences},
  2020.

\bibitem{das2019evaluating}
S.~Das, B.~Wang, Z.~Tingle, and L.~J. Camp.
\newblock Evaluating user perception of multi-factor authentication: A
  systematic review.
\newblock {\em arXiv preprint arXiv:1908.05901}, 2019.

\bibitem{dhamija2006phishing}
R.~Dhamija, J.~D. Tygar, and M.~Hearst.
\newblock {Why Phishing Works}.
\newblock In {\em SIGCHI Conference on Human Factors in Computing Systems},
  pages 581--590, 2006.

\bibitem{fette2007learning}
I.~Fette, N.~Sadeh, and A.~Tomasic.
\newblock {Learning to Detect Phishing Emails}.
\newblock In {\em 16th International Conference on World Wide Web}, pages
  649--656, 2007.

\bibitem{friedrichs2008threat}
O.~Friedrichs, M.~Jakobsson, and C.~Soghoian.
\newblock {The Threat of Political Phishing}.
\newblock In {\em 2nd International Symposium on Human Aspects of Information
  Security \& Assurance}, 2008.

\bibitem{garg2012risk}
V.~Garg, L.~Lorenzen-Huber, L.~J. Camp, and K.~Connelly.
\newblock Risk communication design for older adults.
\newblock In {\em ISARC. Proceedings of the International Symposium on
  Automation and Robotics in Construction}, volume~29, page~1. IAARC
  Publications, 2012.

\bibitem{girden1992anova}
E.~R. Girden.
\newblock {\em {ANOVA: Repeated Measures}}.
\newblock Number~84. Sage Publications Sage CA: Los Angeles, CA, 1992.

\bibitem{hadnagy2010social}
C.~Hadnagy.
\newblock {\em {Social Engineering: The Art of Human Hacking}}.
\newblock John Wiley \& Sons, 2010.

\bibitem{harbach2014using}
M.~Harbach, M.~Hettig, S.~Weber, and M.~Smith.
\newblock {Using Personal Examples to Improve Risk Communication for Security
  \& Privacy Decisions}.
\newblock In {\em SIGCHI Conference on Human Factors in Computing Systems},
  pages 2647--2656, 2014.

\bibitem{hatfield2018social}
J.~M. Hatfield.
\newblock {Social Engineering in Cybersecurity: The Evolution of a Concept}.
\newblock {\em Computers \& Security}, 73:102--113, 2018.

\bibitem{karakasiliotis2006assessing}
A.~Karakasiliotis, S.~Furnell, and M.~Papadaki.
\newblock {Assessing End-User Awareness of Social Engineering and Phishing}.
\newblock In {\em 7th Australian Information Warfare and Security Conference},
  pages 60--72. School of Computer and Information Science, Edith Cowan
  University, Perth, 2006.

\bibitem{kumaraguru2009school}
P.~Kumaraguru, J.~Cranshaw, A.~Acquisti, L.~Cranor, J.~Hong, M.~A. Blair, and
  T.~Pham.
\newblock {School of Phish: A Real-World Evaluation of Anti-Phishing Training}.
\newblock In {\em 5th Symposium on Usable Privacy and Security (SOUPS)}, pages
  1--12, 2009.

\bibitem{lastdrager2017effective}
E.~Lastdrager, I.~C. Gallardo, P.~Hartel, and M.~Junger.
\newblock How effective is anti-phishing training for children?
\newblock In {\em Thirteenth Symposium on Usable Privacy and Security
  ($\{$SOUPS$\}$ 2017)}, pages 229--239, 2017.

\bibitem{martin2018signal}
J.~Martin, C.~Dub{\'e}, and M.~D. Coovert.
\newblock {Signal Detection Theory (SDT) Is Effective for Modeling User
  Behavior Toward Phishing and Spear-Phishing Attacks}.
\newblock {\em Human Factors}, 60(8):1179--1191, 2018.

\bibitem{maurer2011using}
M.-E. Maurer, A.~De~Luca, and S.~Kempe.
\newblock {Using Data Type Based Security Alert Dialogs to Raise Online
  Security Awareness}.
\newblock In {\em 7th Symposium on Usable Privacy and Security (SOUPS)}, pages
  1--13, 2011.

\bibitem{nicholson2020investigating}
J.~Nicholson, Y.~Javed, M.~Dixon, L.~Coventry, O.~Dele-Ajayi, and P.~Anderson.
\newblock Investigating teenagers’ ability to detect phishing messages.
\newblock In {\em EuroUSEC 2020: The 5th European Workshop on Usable Security}.
  IEEE, 2020.

\bibitem{nikou2019impact}
S.~Nikou, M.~Br{\"a}nnback, and G.~Wid{\'e}n.
\newblock {The Impact of Digitalization on Literacy: Digital Immigrants vs.
  Digital Natives}.
\newblock In {\em 27th European Conference on Information Systems}, pages
  1--15. ECIS, 2019.

\bibitem{ometov2018multi}
A.~Ometov, S.~Bezzateev, N.~M{\"a}kitalo, S.~Andreev, T.~Mikkonen, and
  Y.~Koucheryavy.
\newblock {Multi-Factor Authentication: A Survey}.
\newblock {\em Cryptography}, 2(1):1--31, 2018.

\bibitem{pattinson2012some}
M.~Pattinson, C.~Jerram, K.~Parsons, A.~McCormac, and M.~Butavicius.
\newblock {Why do Some People Manage Phishing E-Mails Better than Others?}
\newblock {\em Information Management \& Computer Security}, 20(1):18--28,
  2012.

\bibitem{prakash2010phishnet}
P.~Prakash, M.~Kumar, R.~R. Kompella, and M.~Gupta.
\newblock {Phishnet: Predictive Blacklisting to Detect Phishing Attacks}.
\newblock In {\em 29th IEEE Conference on Computer Communications}, pages 1--5.
  IEEE, 2010.

\bibitem{phishing2018phooled}
P.-P. A.~E. Program.
\newblock {\em Phishing: Don't be Phooled!}, 2018 (accessed June 29, 2020).
\newblock
  \url{"https://www.dhs.gov/sites/default/files/publications/2018_AEP_Vulnerabilities_of_Healthcare_IT_Systems.pdf}.

\bibitem{rajivan2018creative}
P.~Rajivan and C.~Gonzalez.
\newblock {Creative Persuasion: A Study on Adversarial Behaviors and Strategies
  in Phishing Attacks}.
\newblock {\em Frontiers in Psychology}, 9, 2018.

\bibitem{sheng2010falls}
S.~Sheng, M.~Holbrook, P.~Kumaraguru, L.~F. Cranor, and J.~Downs.
\newblock {Who Falls for Phish? A Demographic Analysis of Phishing
  Susceptibility and Effectiveness of Interventions}.
\newblock In {\em SIGCHI Conference on Human Factors in Computing Systems},
  pages 373--382, 2010.

\bibitem{wu2006security}
M.~Wu, R.~C. Miller, and S.~L. Garfinkel.
\newblock {Do Security Toolbars Actually Prevent Phishing Attacks?}
\newblock In {\em SIGCHI Conference on Human Factors in Computing Systems},
  pages 601--610, 2006.

\bibitem{xiang2011cantina+}
G.~Xiang, J.~Hong, C.~P. Rose, and L.~Cranor.
\newblock {Cantina+ A Feature-Rich Machine Learning Framework for Detecting
  Phishing Web Sites}.
\newblock {\em ACM Transactions on Information and System Security (TISSEC)},
  14(2):1--28, 2011.

\end{thebibliography}
\endgroup
\par}
\end{document}